\documentclass[journal]{IEEEtran}

\ifCLASSINFOpdf
\else
   \usepackage[dvips]{graphicx}
\fi
\usepackage{url}

\usepackage{graphicx}

\usepackage{multirow}
\usepackage{xcolor}
\usepackage{soul}
\usepackage{amsfonts} 
\usepackage{cite}
\usepackage{algorithm}
\usepackage{algpseudocode}

\usepackage{hyperref}       
\usepackage{booktabs}       
\begin{document}
\title{Discriminatory and orthogonal feature learning for noise robust keyword spotting}

\author{Donghyeon Kim, Kyungdeuk Ko, \IEEEmembership{Student Member, IEEE}, David K. Han  and Hanseok Ko, \IEEEmembership{Senior Member, IEEE}

\thanks{This work was supported by Korea Environment Industry Technology Institute(KEITI) through Exotic Invasive Species Management Program, funded by Korea Ministry of Environment(MOE)
(2021002280004)
}
\thanks{D. Kim, K. Ko and H. Ko are with the School of Electrical Engineering,
Korea University, Seoul 02841, South Korea (e-mail: kis6470@korea.ac.kr;kdko@korea.ac.kr;hsko@korea.ac.kr).}
\thanks{D. K. Han is with Department of Electrical and Computer Engineering, Drexel University, Philadelphia, USA (e-mail: dkh42@drexel.edu).}
}

\markboth{Journal of \LaTeX\ Class Files, Vol. 14, No. 8, August 2015}
{Shell \MakeLowercase{\textit{et al.}}: Bare Demo of IEEEtran.cls for IEEE Journals}
\maketitle

\begin{abstract}
Keyword Spotting (KWS) is an essential component in a smart device for alerting the system when a user prompts it with a command. As these devices are typically constrained by computational and energy resources, the KWS model should be designed with a small footprint. In our previous work, we developed lightweight dynamic filters which extract a robust feature map within a noisy environment. The learning variables of the dynamic filter are jointly optimized with KWS weights by using Cross-Entropy (CE) loss. CE loss alone, however, is not sufficient for high performance when the SNR is low. In order to train the network for more robust performance in noisy environments, we introduce the LOw Variant Orthogonal (LOVO) loss. The LOVO loss is composed of a triplet loss applied on the output of the dynamic filter, a spectral norm-based orthogonal loss, and an inner class distance loss applied in the KWS model. These losses are particularly useful in encouraging the network to extract discriminatory features in unseen noise environments. 
\end{abstract}


\IEEEpeerreviewmaketitle

\section{Introduction}
\IEEEPARstart{I}{n} audio-based deep learning applications, mitigating noise disturbance in real audio streams is challenging work and several studies have been conducted to combat noise issues. A Denoising AutoEncoders (DAE) have shown reasonable performances by reducing the distance between the clean audio and the output of the decoder model\cite{badi2020correlation,zhang2018multi,sun2015unseen}.
Other proposed methods include: Generative Adversarial Network(GAN) \cite{pascual2017segan}, Griffin-Lim \cite{masuyama2019deep}, evaluation metric learning \cite{fu2018end} and etc. Instead of reducing noise directly, a feature enhancement by the classification loss \cite{shon2019voiceid,kim2020dual} has been proposed. These methods are typically used at the front-end of the main task model and they are jointly optimized through the classification loss function. However, since these front-end methods require a high degree of computation power, they might not be appropriate with limited resources such as Keyword Spotting (KWS) to detect trigger speech \cite{michaely2017keyword}. As the KWS is applied to the streaming environment, its network should be minimally designed to improve the model efficiency. For this reason, the number of model parameters and FLOPS are key elements to evaluate the KWS performance and several deep learning methods have been developed. For the memory-efficient Convolutional Neural Network (CNN), a Depthwise Separable Convolution (DSConv) \cite{zhang2017hello} and a temporal 1D CNN structure \cite{choi2019temporal,li2020small} show reasonable performance with relatively low computational costs. Also, raw audio-based methods \cite{kumatani2017direct,mittermaier2020small} would save the computational resource for the acoustic feature extraction by keeping the KWS performance. In addition, Neural Architecture Search (NAS) \cite{mo2020neural,zhang2021autokws}, far-field data augmentation \cite{gao2020towards} and joint model with speaker verification \cite{rikhye2021personalized} have been used to improve the KWS performance.

In our earlier work \cite{kim2021lightweight}, we confirmed that applying a lightweight dynamic filter to the front-end of the temporal CNN would improve the performance of the KWS in a noisy environment. However, as the networks are trained by the Cross-Entropy (CE) loss, it may be difficult to extract the robust feature. 
In order to extract the robust feature, the networks are trained to find robust domains, while CE loss aims to find separable features for the classification task. 
For the robust feature against noisy environments, we aim to design the network to extract embedded features with the following three considerations: Inter-class variance, Intra-class variance, and Inter-class orthogonality. 

These issues are covered by metric learning approaches. Conventional metric learning methods \cite{wen2016discriminative,schroff2015facenet,cogswell2015reducing}, sample mining strategy \cite{wang2019multi} and circular decision boundary \cite{sun2020circle} are applied to extract discriminative embedding.
While these methods delivered improved performance, the pairwise cosine or Euclidean distance based methods have limitations.
The embedding for methods that use pairwise metric learning \cite{schroff2015facenet,wang2019multi,sun2020circle} may be biased to dominant classes that are easy to train or have large training samples \cite{sinha2020class}.
To combat this issue, we propose LOw Variability Orthogonal (LOVO) loss to enhance the robustness against noise environments. To this end, we divide our previously studied model into the dynamic filter and the classifier for two-way metric learning.
In the output of the dynamic filter, we utilize pairwise triplet loss \cite{schroff2015facenet} to improve the feature separability. We then apply class centroid-based metric learning to the embedding vector of the classifier by considering the class-wise variance and class-wise orthogonality. The Euclidean distance is used in the class-wise variance to increase the inter-class distance and decrease the intra-class distance. For the class-wise orthogonality, a Spectral Norm (SN) \cite{bansal2018can} based loss is used. As the metric learning is performed on the class centroids, the network is less biased to the dominant classes and shows robust performance in noisy environments. 

The KWS experiments are carried out on Speech
Command dataset \cite{warden2018speech}. 
We compare our method with the small footprint KWS models and various metric learning methods in unseen noise environments. 
Our proposed method archives 3\% performance improvement in the 0dB condition over our previous work. Additionally, we confirm that our method shows robust performance in the low Signal-to-Noise Ratio (SNR) conditions over various metric learning methods. 
\section{LOVO embedding vector}
To achieve our goals, we consider two models for metric learning, and the pipeline is shown in Fig \uppercase\expandafter{\romannumeral1}.\\
{\bf Dynamic embedding model.} The aim of the dynamic embedding model is to extract embedding vectors that are strongly connected with the dynamic filter. The output of the dynamic filter is fed to the dynamic embedding model and the triplet loss is computed. The computed loss is used to update the weights of the dynamic filter-related model and the dynamic embedding model would not be computed in the test phase.\\
{\bf KWS model.} KWS model is a classifier for the KWS implementation. The output of the dynamic filter is fed to the classifier and the CE is computed. Metric learning is performed on an embedding vector of the KWS model.\\
These loss functions would be used in training the network to extract LOVO embedding vectors. Since the loss functions are only computed during the training process, there is no additional cost to the KWS implementation.
\subsection{Dynamic filter embedding}
\begin{figure}[t]
     \centering
     \includegraphics[scale=0.24]{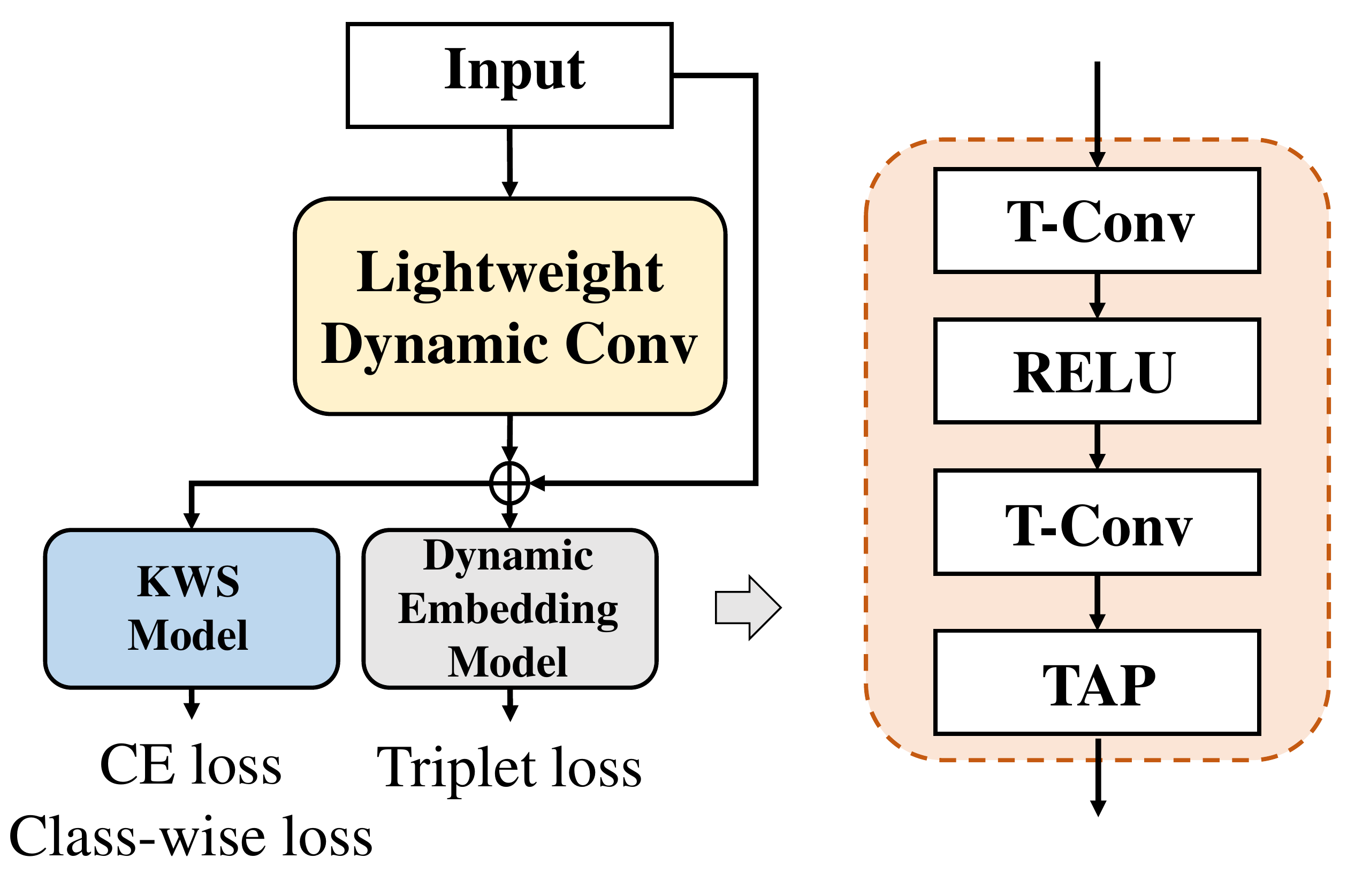}
     \caption{Pipeline of the proposed method. T-Conv denotes the temporal convolution layer.}
    \label{fig:three graphs}
\end{figure}
To enhance the performance of the dynamic filter directly, we apply triplet loss \cite{schroff2015facenet} based metric learning to the output of the dynamic model. The dynamic model is based on two layers of temporal convolutions with a Rectified Linear Unit (ReLU) and a Temporal Averaging Pooling (TAP) to produce the embedding vector. The details of the process are as follows:
\begin{equation}
  H_i = TAP(Conv(max(0,(Conv(x^{'}_i, \mathbf{w_1})))), \mathbf{w_2}),
  \label{eq8}
\end{equation}
where $\mathbf{w_1}\in R^{[K, F, F_{df}]}$ and $\mathbf{w_2}\in R^{[K, F_{df}, F_{dim}]}$ denote learning variables for the metric learning. The TAP denotes feature averaging to the temporal dimension. From this process, $F_{dim}$ dimension of the embedding vector ($H_i$) is produced and the triplet loss is computed as follows: 
\begin{equation}
  L_{M} = \sum_{i=1}^{M}\sum_{j=1}^{M} y_i *\vert \vert{H_{i}-H_{j}}\vert \vert^2_2+\alpha,
  \label{eq2}
\end{equation}
where $y_i$ denotes the label of the metric learning model which returns $1$ when $H_i$ and $H_j$ are of the same class and $-1$ when they are of a different class. $\alpha$ and $M$ denote the margin of the loss and the size of the mini-batch, respectively. As such, the loss function minimizes the Euclidean distance of the intra-class embedding while it maximizes the Euclidean distance of the inter-class embedding. Our aim in the process is to train the weights of the dynamic filter to extract task-relevant salient features among the classes.
\subsection{Keyword embedding}
The metric learning in keyword embedding ($E$) aims to
form tighter clusters within the same class while encouraging embedding centroid of each class to be orthogonal to each other. For the keyword model, we utilize TENet12 \cite{li2020small} and the embedding vectors are obtained from the output of the final CNN layer. For tighter cluster formation, we build class-wise subsets where only the same class embedding vectors are assembled from the mini-batch unit. Then, the distances between the embedding vectors and their centroids are minimized by the following equation:   
\begin{equation}
  L_{I} = \frac{1}{C} \sum_{k=1}^{C}\sum_{i=1}^{M_k} \vert \vert E^k_{i}-\overline{E^k}\vert \vert^2_2,
  \label{eq3}
\end{equation}
where $E^k_{i} \in \mathbb{R}^{[D,M_k]}$ denotes the $k^{th}$ class subset which has $D$ dimension and $M_k$ is the sample size. C and $\overline{E^k} \in \mathbb{R}^{[D,1]}$ denote the number of class and class centroid respectively.
Additionally, we implement an inter-class orthogonal loss by forming a combined matrix $E^{'}$ by following equation:
\begin{equation}
  E^{'} = [\overline{E^1},\overline{E^2},...,\overline{E^c}],
  \label{eq4}
\end{equation}
where $E^{'} \in \mathbb{R}^{[D,C]}$. 
We compute the Euclidean distance among the class centroids and then, concatenate them to form a class-wise distance matrix as follows:
\begin{equation}
  M_{DM} = [||E^{'}-\overline{E^1}||^2_2,||E^{'}-\overline{E^2}||^2_2,...,||E^{'}-\overline{E^c}||^2_2],
  \label{eq5}
\end{equation}
where $M_{DM} \in \mathbb{R}^{[C,C]}$ whose diagonal elements are zero. We define $M_{DM}$ as a loss function related to inter-class distance. To enable class-wise orthogonality of the centroids, we utilize the covariance of $E^{'}$ as follows:
\begin{equation}
  M_{IM}=\frac{1}{C-1}(E^{'}-\overline{E^{'}})^T(E^{'}-\overline{E^{'}}),
  \label{eq6}
\end{equation}
where $\overline{E^{'}} \in \mathbb{R}^{[D,1]}$ denotes the combined global mean vector of all the centroids. $M_{IM} \in \mathbb{R}^{[C,C]}$ is used to compute the orthogonality loss. As we aim to increase the distance of the inter-class centroids and to enforce class-wise orthogonality among the centroids, the Frobenius norm of $M_{DM}$ should be maximized while the Frobenius norm of $M_{IM}-I$ should be minimized. As these loss functions might compete with one another in some parts of the training \cite{sener2018multi}, we combine the two-loss functions as one. Since inter-class orthogonality would also encourage inter-class cluster distances, the influence of $M_{IM}-I$ is extended to a larger extent compared to that of $M_{DM}$ by applying an exponential function to limit the range of $M_{DM}$. Secondly, we utilize the SN \cite{miyato2018spectral,bansal2018can} as the object for minimizing the loss. Instead of directly applying the loss, minimizing the SN of the loss function shows more stable and improved performance over the Euclidean distance. The equation of the SN is as follows:
\begin{equation}
  L_{O} = SN(p(M_{IM})+e^{-M_{DM}}-I),
\end{equation}
where $SN(\cdot)$ denotes the spectral norm. As orthogonality is related to the non-diagonal elements in $M_{IM}$, we utilize a matrix filter ($p(\cdot)$) to maps diagonal elements to zero. As a result, the maximum singular value in equation 7 would be considered as the loss function and it is computed by using a power iteration method. It's been found that the optimal number of iterations was 10.
For the model training, loss functions are jointly optimized with CE loss as follows:
\begin{equation}
  L_{Total} = L_{CE}+\lambda_1 L_{M}+\lambda_2 L_{I}+ \lambda_3 L_{O},
  \label{eq7}
\end{equation}
where $L_{CE}$ denotes CE loss between the labels and predictions of the classification model. The total loss ($L_{total}$) is computed by adding $L_{CE}$, $L_{M}$, $L_{I}$ and $L_{O}$ with different scales of $\lambda_*$. We consider LOVO loss as a combination of the metric learning losses. It has been found empirically that the effect of $L_{I}$ and $L_{O}$ during the model training is dominant over other learning losses and it might disturb the learning of $L_{CE}$. So, we set $\alpha=1$, $\lambda_1=0.25$, $\lambda_2=0.01$ and $\lambda_3=0.01$ as the weights and coefficient for the learning balance.
\section{Keyword Spotting}
\subsection{Experimental Setup}
we used Speech Command v1 \cite{warden2018speech} to evaluate our method. The Speech Command dataset contains 30 keywords and 10 keywords with two extra classes (unknown and silence) were utilized for the model training. We utilized 80\% of the dataset for the model training, 10\% for validation, and the remaining 10\% for the test. By following the DB guideline of the dataset, we injected background noise (mike noise) and performed random time-shifting. For evaluating robustness over unseen noise, we utilized DCASE \cite{mesaros2019acoustic}, Urbansound8K \cite{salamon2014dataset} and WHAM \cite{Wichern2019WHAM} datasets as the unseen noise dataset. They contain the background noise of urban areas. For generating noisy data, we randomly selected audio samples from the noise data and mixed them with the original test data. Five different Signal-to-Noise Ratios (SNR) levels [20dB, 15dB, 10dB, 5dB and 0dB] were applied.

In the training process, we used a batch size of 100, a learning step of 30K, and ADAM optimizer \cite{kingma2014adam} with a 0.001 initial learning rate. Every 10K step, the learning rate was decremented by a factor of 0.1. We used 40 dimensions of MFCC as input T-F features. It was constructed with 30ms of windows with 10ms overlap and 64 Mel filters. In the dynamic embedding model, the temporal convolutions was computed by $w_1\in\mathbb{R}^{[9,40, 40]}$ and $w_2\in\mathbb{R}^{[9,40, 128]}$ size of weights with stride of 2. We utilized 32 dimensions of the embedding vector for the LOVO loss, which is the output of the global averaging pooling layer in the TENet12 model. 
\subsection{Baselines}
We implement our proposed loss functions on the LDy-TENet12 \cite{kim2021lightweight} and compared its performance with the following baseline models.\\
{\bf TCNet.} TCNet14 \cite{choi2019temporal} contains blocks of temporal convolutions and a skip-connection. TCNet14 is composed of 6 of these convolution and skip-connection blocks and 1 FC layer. Each convolution block has two temporal convolution layers.\\
{\bf TENet.} TENet12 \cite{li2020small} uses depth-wise and point-wise temporal convolutions. TENet12 contains 12 convolution blocks with 1 FC layer. Every convolution block has 32 output channels.\\
{\bf Neural Architecture Search.} NAS is a network designing method of using search costs (FLOPS, memory, accuracy, etc.). We compare our method with several NAS methods including Differentiable Architecture Search (DARTS). Please see details of the model in \cite{mo2020neural,zhang2021autokws}.\\
{\bf SincConv.} SincConv \cite{mittermaier2020small} is a raw audio-based model with a learnable sinc function \cite{ravanelli2018speaker}. By utilizing two sinc functions, frequency bandpass filters are obtained, and they are used to extract acoustic features of the raw audio. Then DSConv and Group DSConv are employed as a classifier.\\
{\bf Lightweight convolution.} Lightweight Convolution \cite{wu2019pay} (LConv.) block is designed to perform a group separable convolution with a weight normalization. It has two linear layers, Gated Linear Unit (GLU), and the group separable convolution. We use the single LConv block in the front end of the TENet12 model. In the first linear layer, frequency dimension of the T-F features is increased by 80 and the GLU is applied. Then, the separable convolution with $H=10$ and the other linear layer is conducted. Additionally, instead of the static weights, we utilize a single linear layer to produce the weights for the separable convolution (DyConv).\\
{\bf LDy-TENet.} LDy-TENet12 \cite{kim2021lightweight} uses a lightweight dynamic convolution on the TENet12 \cite{li2020small} based model. $3\times3$ CNN kernel ($k=9$) is employed to perform the PDF and the dynamic convolution process. In the IDF, the first and second FC follow $40 \times 40$ and $40 \times k$ dimensions, respectively.
\subsection{Result discussion}
\begin{table}[t]
    \centering
    \scriptsize
  \label{tab:1}
  \begin{tabular}{|c|c|c|c|c|}
    \hline
    \textbf{Model}& \textbf{Param.}& \textbf{FLOPS.}& \textbf{Avg. Acc}& \textbf{Best}\\
    \hline
    TCNet14\cite{choi2019temporal}& 305K&8.26M& - &96.6 \\
    \hline
    TENet6-n\cite{li2020small}& 17K&1.26M& - &96.0 \\
    TENet12-n\cite{li2020small}& 31K&1.97M& - &96.3 \\
    TENet6\cite{li2020small}& 54K&3.95M& - &96.4 \\
    TENet12\cite{li2020small}& 100K&6.42M& - &96.6 \\
    \hline
    NAS2\cite{mo2020neural}& 886K&-& - &97.2 \\
    Random\cite{zhang2021autokws}& 196K&8.8M& 96.58 &96.8 \\
    DARTS\cite{zhang2021autokws}& 93K&4.9M& 96.63 &96.9 \\
    F-DARTS\cite{zhang2021autokws}& 188K&10.6M& 96.70 &96.9 \\
    N-DARTS\cite{zhang2021autokws}& 109K&6.3M& 96.79 &97.2 \\
    \hline
    SincConv-DS\cite{mittermaier2020small}& 122K&-& - &96.6 \\
    SincConv-GDS\cite{mittermaier2020small}& 62K&-& - &96.4 \\
    \hline
    LConv\cite{wu2019pay}& 105K&7.40M& 96.88 &97.0 \\
    DyConv\cite{wu2019pay}& 107K&7.69M& 96.89 &97.1 \\
    \hline
    LDy-TENet6-n\cite{kim2021lightweight}& 19K&1.48M& 96.48 &97.0 \\
    LDy-TENet12-n\cite{kim2021lightweight}& 33K&2.19M& 96.69 &96.9 \\
    LDy-TENet6\cite{kim2021lightweight}& 56K&4.17M& 96.77 &96.9 \\
    LDy-TENet12\cite{kim2021lightweight}& 102K&6.64M& 96.95 &97.1 \\
    \hline
    LDy$_{L_M}$-TENet12& 102K&6.64M& 97.10 &97.3 \\
    LDy$_{L_{MIO}}$-TENet12& 102K&6.64M& 97.00 &97.2 \\
    \hline
  \end{tabular}
    \caption{Comparison with lightweight models on Speech
Command v1. Param. and FLOPS. denote Model parameters
and computational cost respectively}
\end{table}

\begin{table*}[t]
\scriptsize
  \label{tab:2}
  \centering
\begin{tabular}{|c|c|cccccccccccc|}
    \hline
\multicolumn{1}{|c|}{\multirow{2}{*}{\textbf{Noise}}} & \multicolumn{1}{c|}{\multirow{2}{*}{{\begin{tabular}[c]{@{}c@{}}\textbf{SNR}\\ (dB)\end{tabular}}}} & \multicolumn{12}{c|}{\textbf{Model}}\\\cline{3-14}
\multicolumn{1}{|c|}{}&\multicolumn{1}{c|}{}&\textbf{$L_{MIO}$}&\textbf{$L_{MO}$}&\textbf{$L_{MI}$}&\textbf{$L_{IO}$}&\textbf{$L_{O}$}&\textbf{$L_{I}$}&\textbf{$L_{M}$}&\textbf{LDy}&\textbf{Dconv} &\textbf{Lconv} & \textbf{TENet}& \textbf{TCNet}\\    \hline
\multicolumn{2}{|c|}{\textbf{CLEAN}}&97.00 &97.04 &97.12 &97.07 & 96.83&\textbf{97.12} &97.10 &96.95&  96.89 &  96.88 &96.60&96.60\\    \hline
\multirow{5}{*}{{\begin{tabular}[c]{@{}c@{}}\textbf{DCASE}\\\cite{mesaros2019acoustic}\end{tabular}}}  
                                &\textbf{20}&96.79 &96.79 &96.77 &96.81 &96.61 & 96.80&\textbf{96.88}&96.58& 96.55 & 96.62 &96.28&95.85\\
						        &\textbf{15}&96.54 &96.61 &96.63 &96.62 &96.43 & 96.60&\textbf{96.71}&96.44& 96.35 & 96.37 &96.23&95.59\\
					            &\textbf{10}&95.49 &95.48 &\textbf{95.57} &95.51 &95.29 & 95.48&95.48&95.10& 95.01 & 95.05 &94.94&94.25\\
                                &\textbf{5}&\textbf{93.74} &93.47 &93.61 &93.57 &93.24 & 93.52&93.33&93.00& 92.27 & 92.80 &92.87&91.79\\
					            &\textbf{0}&\textbf{89.71} &89.16 &89.46 &89.43 &88.97 & 89.32&88.90&88.41& 87.51 & 87.67 &88.14&86.01\\    \hline
\multirow{5}{*}{{\begin{tabular}[c]{@{}c@{}}\textbf{Urban}\\\cite{salamon2014dataset}\end{tabular}}}   
                                &\textbf{20}&96.10 &\textbf{96.17} &96.16 &96.21 &96.04 & 96.07&\textbf{96.17}&95.94& 95.81 & 95.84 &95.72&95.16\\
						        &\textbf{15}&95.16 &95.12 &95.18 &\textbf{95.25} &95.17 & 95.20&95.07&95.09& 94.69 &  94.49 &94.69&93.61\\
					            &\textbf{10}&\textbf{93.31} &93.12 &\textbf{93.31} &93.12 &92.94 & 93.02&93.16&92.82& 92.03 &  92.25 &92.17&90.77\\
                                &\textbf{5}&\textbf{90.30} &89.91 &90.19 &90.06 &89.61 & 89.83&89.53&89.19& 87.80 & 87.60 &87.97&86.38\\
					            &\textbf{0}&\textbf{80.92} &80.19 &80.38 &80.57 &79.47 & 79.91&79.87&78.55& 76.20 & 76.26 &77.54&74.20\\    \hline
\multirow{5}{*}{{\begin{tabular}[c]{@{}c@{}}\textbf{WHAM}\\\cite{Wichern2019WHAM}\end{tabular}}}   
                                &\textbf{20}&96.27 & 96.34&\textbf{96.32} &96.26 &96.12 &96.24 &96.29&96.09& 95.96 & 96.02 &95.75&95.43\\
						        &\textbf{15}&95.70 & 95.63&\textbf{95.79} &95.69 &95.57 &95.68 &95.62&95.47& 96.12 & 95.23 &95.12&94.44\\
					            &\textbf{10}&\textbf{93.60} & 93.26&93.20 &93.39 &93.02 &93.50 &93.01&92.91& 92,39 & 92.69 &92.67&91.46\\
                                &\textbf{5}&\textbf{89.43} & 88.89&89.16 &89.25 &88.42 &89.11 &88.94&88.19& 87.36 & 87.83 &87.88&85.72\\
					            &\textbf{0}&\textbf{78.95} &77.26 &77.87 &78.29 &76.29 &77.62 &76.93&75.92& 74.52 & 75.18 &75.19&73.15\\    \hline
\multicolumn{2}{|c|}{\textbf{AVG. Accuracy}}& \textbf{92.63}& 92.15&92.29 &92.32 &91.88 &92.19 &92.09&91.66& 90.64 &  90.79 &91.23&90.03\\
    \hline

\end{tabular}
  \caption{Comparison with unseen noise environment. Experiments are performed on LDy-TENet12, lightweight convolutions, TENet12, and TCNEt14 models. AVG. Accuracy denote mean accuracy of the clean and noisy environments}
\end{table*}

For accurate model evaluation, 8 repeated experiments are carried out. Our proposed method is only applied to the LDy-TENet12 model (LDy). We compare our method with the small footprint KWS models and various metric learning models.\\
{\bf Small footprint KWS.} The results in Tables \uppercase\expandafter{\romannumeral1} and \uppercase\expandafter{\romannumeral2} show the KWS performance based on the small footprint KWS models in the original test set and noise environments, respectively. We confirm that $L_M$ has the best performance in a clean environment. It means that metric learning in the output of the dynamic filter would help to extract the discriminative feature for the KWS. On the other hand, $L_{MIO}$ shows superior performance, particularly in the low SNR (5dB and 0dB) conditions. Applying $L_{O}$ and $L_{M}$ would also show robust performance in a noisy environment. Among the metric loss functions, we observe that $L_{I}$ loss results in higher performance over ($L_{O}$ and $L_{M}$) losses, and jointly applying two losses ($L_{IO}$) delivers significantly robust performance in high noise environments. As $L_{I}$ considers inter-class variance and $L_{O}$ considers inter-class distance and orthogonality, applying $L_{O}$ alone would not ensure the distance margin among the classes is truly enhanced. For this reason, jointly utilizing the two-loss function ($L_{IO}$) would mitigate this uncertainty.\\
{\bf Comparison with metric learning.} Table \uppercase\expandafter{\romannumeral3} shows a comparison of the LOVO loss and other metric learning methods \cite{wen2016discriminative,schroff2015facenet,cogswell2015reducing,wang2019multi,sun2020circle} with the same KWS model (LDy-TENet 12). As the other metric learning-based methods are employed solely for the KWS embedding vector, we only implemented $L_{IO}$ in our model for a fair comparison and excluded $L_{M}$.
The results indicate that the pairwise learning methods \cite{schroff2015facenet,wang2019multi,sun2020circle} perform better at high SNR levels, while other methods, including ours, which is not pairwise, show robust performance at low SNR levels. As the pairwise distance computes all possible pairs in the mini-batch unit, it shows improved performance in the clean and high SNR noisy environments which are similar to the clean data while they show degraded performance as the network is biased to the dominant classes. In contrast, non-pairwise methods show little performance degradation in the clean environment, and it takes robustness in the low SNR conditions.
Particularly, as our method would map the input data to the orthogonal class centroids, it shows robust performance over the other methods. Although the performance improvement between $L_{IO}$ and \cite{cogswell2015reducing} is not significant, $L_{MIO}$ which is our proposed method produces a $1.2\%$ improvement in the WHAM 0dB condition. By considering the computational power of the classifier and the performance in a clean environment (almost 97\%), our method shows reasonable performance in noisy environment. 

\begin{table}[t]
\scriptsize
  \label{tab:3}
  \centering
\begin{tabular}{|c|c|cccccc|}
    \hline
\multicolumn{1}{|c|}{\multirow{2}{*}{\textbf{Noise}}} & \multicolumn{1}{c|}{\multirow{2}{*}{{\begin{tabular}[c]{@{}c@{}}\textbf{SNR}\\ (dB)\end{tabular}}}} & \multicolumn{6}{c|}{\textbf{Method}}\\ \cline{3-8}
\multicolumn{1}{|c|}{}&\multicolumn{1}{c|}{}&\textbf{$L_{IO}$}&\cite{sun2020circle} &\cite{wang2019multi}&\cite{schroff2015facenet}& \cite{wen2016discriminative}& \cite{cogswell2015reducing}\\    \hline
\multicolumn{2}{|c|}{\textbf{CLEAN}}&97.07&97.09 &96.99&\textbf{97.10} &96.97 &96.99 \\    \hline
\multirow{5}{*}{{\begin{tabular}[c]{@{}c@{}}\textbf{DCASE}\\\cite{mesaros2019acoustic}\end{tabular}}}  
                                &\textbf{20}&96.81& \textbf{96.95}&96.69 &96.84&96.64&96.79\\
						        &\textbf{15}&96.62& \textbf{96.71}&96.45 &96.67&96.47&96.55\\
					            &\textbf{10}&95.51& \textbf{95.67}&95.43 &95.64&95.26&95.62\\
                                &\textbf{5}&\textbf{93.57}& 93.55&93.19 &93.47&93.35&93.55\\
					            &\textbf{0}&\textbf{89.43}& 89.07&88.88 &88.97&89.17&89.29\\    \hline
\multirow{5}{*}{{\begin{tabular}[c]{@{}c@{}}\textbf{Urban}\\\cite{salamon2014dataset}\end{tabular}}}   
                                &\textbf{20}&96.21& \textbf{96.32}&96.15 &96.26&95.97&96.12\\
						        &\textbf{15}&95.25&\textbf{95.40} &95.14 &95.22&95.08&95.23\\
					            &\textbf{10}&93.12&93.12 &93.01 &93.22&92.87&\textbf{93.34}\\
                                &\textbf{5}&\textbf{90.06}&89.57 &89.37 &89.69&89.59&89.78\\
					            &\textbf{0}&\textbf{80.57}&79.07 &79.40 &79.27&79.67&80.14\\    \hline
\multirow{5}{*}{{\begin{tabular}[c]{@{}c@{}}\textbf{WHAM}\\\cite{Wichern2019WHAM}\end{tabular}}}   
                                &\textbf{20}&96.26&96.39 &96.11 &\textbf{96.41}&96.16&96.25\\
						        &\textbf{15}&95.69&\textbf{95.75} &95.67 &\textbf{95.75}&95.50&95.61\\
					            &\textbf{10}&\textbf{93.39}&93.16 & 93.09&93.34&93.25&\textbf{93.39}\\
                                &\textbf{5}&\textbf{89.25}&88.83 &88.58 &88.64&89.00&89.15\\
					            &\textbf{0}&\textbf{78.29}&76.56 &76.78 &76.34&77.06&77.72\\    \hline
\multicolumn{2}{|c|}{\textbf{Total-AVG.}}&\textbf{92.32}&92.08 &91.93 &92.05&92.00&92.22\\
    \hline
\end{tabular}
  \caption{Comparison with other metric learning methods.}
\end{table}
\section{Conclusion}
In this letter, we proposed the LOVO loss function which aims to extract orthogonal and discriminatory features useful in the KWS. To this end, we applied metric learning for the dynamic filter and the KWS model. Firstly, we used triplet loss to the dynamic filter to enhance inter-class separability and intra-class clustering of the features. Then, the spectral norm-based orthogonal loss and intra-class distance were minimized in the KWS embedding vector. From these processes, the dynamic filter and the KWS model were encouraged to capture salient features to enhance classification performance. The experimental results also showed that our proposed method enhances KWS performance over low SNR unseen noise without additional computational resources.


\bibliographystyle{IEEEtran}
\bibliography{bibfile}

\end{document}